# Gradient Compared $\ell_p$-LMS Algorithms for Sparse System Identification

Yong Feng[1,2], Jiasong Wu[2], Rui Zeng[2], Limin Luo[2], Huazhong Shu[2]

1. School of Biological Science and Medical Engineering, Southeast University, Nanjing 210096, China
E-mail: fengyong@seu.edu.cn

2. Laboratory of Image Science and Technology, the Key Laboratory of Computer Network and Information Integration, Southeast University, Nanjing 210096, China
E-mail: shu.list@seu.edu.cn

**Abstract:** In this paper, we propose two novel *p*-norm penalty least mean square ($\ell_p$-LMS) algorithms as supplements of the conventional $\ell_p$-LMS algorithm established for sparse adaptive filtering recently. A gradient comparator is employed to selectively apply the zero attractor of *p*-norm constraint for only those taps that have the same polarity as that of the gradient of the squared instantaneous error, which leads to the new proposed gradient compared *p*-norm constraint LMS algorithm ($\ell_p$GC-LMS). We explain that the $\ell_p$GC-LMS can achieve lower mean square error than the standard $\ell_p$-LMS algorithm theoretically and experimentally. To further improve the performance of the filter, the $\ell_p$NGC-LMS algorithm is derived using a new gradient comparator which takes the sign-smoothed version of the previous one. The performance of the $\ell_p$NGC-LMS is superior to that of the $\ell_p$GC-LMS in theory and in simulations. Moreover, these two comparators can be easily applied to other norm constraint LMS algorithms to derive some new approaches for sparse adaptive filtering. The numerical simulation results show that the two proposed algorithms achieve better performance than the standard LMS algorithm and $\ell_p$-LMS algorithm in terms of convergence rate and steady-state behavior in sparse system identification settings.

## 1 INTRODUCTION

Sparse systems, whose impulse responses contain many near-zero coefficients and few large ones, are very common in practice. For example, sparse wireless multi-path channels, sparse acoustic echo path, digital TV transmission channels, and so forth. The least mean square (LMS) algorithm [1] is the most widely used technique in applications like system identification (SI) which plays an important role in the application of adaptive filtering. However, the traditional LMS algorithm does not assume any structural information about the system to be identified and thus performs poorly both in terms of steady-state excess mean square error (excess MSE) and convergence rate [2].

Recently there emerges a growing research interest in sparse system identification, for example, sparse LMS algorithms with different norm constraints, which are mainly motivated by research of the least absolute shrinkage and selection operator (LASSO) [3] and compressive sensing (CS) [4]. The family of norm constraint LMS algorithms has become one of the main sparse LMS algorithms in adaptive filtering during the last few years [5].

This work is supported by the National Basic Research Program of China under Grant 2011CB707904, by the NSFC under Grants 61201344, 61271312, 11301074, and by the SRFDP under Grants 20110092110023 and 20120092120036, the Project-sponsored by SRF for ROCS, SEM, and by Natural Science Foundation of Jiangsu Province under Grant BK2012329 and by Qing Lan Project.

Many norm constraint LMS algorithms have been proposed so far, for instance, $\ell_1$-norm penalty LMS ($\ell_1$-LMS) [1, 6], $\ell_0$-norm penalty LMS ($\ell_0$-LMS) [7, 8] and $\ell_p$-norm penalty LMS ($\ell_p$-LMS) [9, 10], where the corresponding $\ell_1$, $\ell_0$ and $\ell_p$ norms are incorporated into the cost function of the standard LMS algorithm respectively, to increase the convergence speed and decrease the MSE as well. However, these algorithms apply the norm constraint of zero attractors to all the weight taps of the unknown system in general, leading to a slowing down of the convergence for those taps who have the same polarity as the gradient of the squared error [11]. Fortunately, it can be improved by using a gradient compactor that we will shown in the later sections.

To kick off the limitation above, we propose a gradient compared *p*-norm penalty LMS algorithm ($\ell_p$GC-LMS) as well as its improved version the $\ell_p$NGC-LMS, as supplements of the conventional $\ell_p$-LMS algorithm. Numerical simulations show that the proposed algorithms achieve better performance than the standard LMS and $\ell_p$-LMS algorithms in sparse system identification settings. In addition, Taheri and Vorobyov have demonstrated that the $\ell_p$-LMS is superior to other norm constraint LMS algorithms in convergence behavior under certain conditions [9].

The paper is organized as follows: In Section II, we propose the $\ell_p$GC-LMS and its extension the $\ell_p$NGC-LMS algorithm for sparse systems in details, after a brief review of the standard LMS and $\ell_p$-LMS algorithms. Then the numerical simulations are given in Section III to investigate the performance of the two proposed algorithms. Finally, the paper is concluded in Section IV.

## 2 PROPOSED ALGORITHMS

Throughout this paper, matrices and vectors are denoted by boldface upper-case letters and boldface lower-case letters, respectively, while variables and constants are in italic lower-case letters. The superscripts $(\cdot)^T$ represents the transpose operators, and $E[\cdot]$ denotes the expectation operator.

### 2.1 Reviewing standard LMS and $\ell_p$-LMS algorithms

Let $y_k$ be the output of an unknown system with an additional noise $n_k$ at time $k$, which can be written as

$$y_k = \mathbf{w}^T \mathbf{x}_k + n_k, \quad (1)$$

where the weight $\mathbf{w}$ of length $N$ is the impulse response of the unknown system, $\mathbf{x}_k$ represents the input vector with covariance $\mathbf{R}$, defined as $\mathbf{x}_k = [x_k, x_{k-1}, \cdots, x_{k-N+1}]^T$, and $n_k$ is a stationary noise with zero mean and variance $\sigma_k^2$.

Given the input $\mathbf{x}_k$ and output $y_k$ in an unknown linear system following the above settings, the LMS algorithm was proposed to estimate the weight vector $\mathbf{w}$ decades ago. The cost function $J_k$ of the standard LMS algorithm is defined as

$$J_k = e_k^2 / 2, \quad (2)$$

where $e_k = y_k - \mathbf{w}_k^T \mathbf{x}_k$ denotes the instantaneous error and $\mathbf{w}_k = [[\mathbf{w}_k]_1, [\mathbf{w}_k]_2, \ldots, [\mathbf{w}_k]_N]^T$ is the estimated weight of the system at time $k$. Note that the "1/2" here is taken just for the convenience of computation. Thus, using gradient descent, the update equation is written as

$$\mathbf{w}_{k+1} = \mathbf{w}_k - \mu \frac{\partial J_k}{\partial \mathbf{w}_k} = \mathbf{w}_k + \mu e_k \mathbf{x}_k, \quad (3)$$

where $\mu$ is the step size such that $0 < \mu < \lambda_{max}^{-1}$ with $\lambda_{max}$ being the maximum eigenvalue of $\mathbf{R}$.

For the sparse system identification in which most of the taps in the weight vector are exactly or nearly zeros, the $\ell_p$-LMS [9] algorithm has been proposed with the new cost function $J_{k,p}$ written as

$$J_{k,p} = e_k^2 / 2 + \gamma \|\mathbf{w}_k\|_p, \quad (4)$$

where the $p$ norm is defined as $\|\mathbf{w}_k\|_p \triangleq \left(\sum_{i=1}^{N} |[\mathbf{w}_k]_i|^p\right)^{1/p}$ with $0 < p < 1$, and $\gamma$ is a constant controlling the trade-off between the convergence speed and estimation error. Thus, the update of the $\ell_p$-LMS is then derived as

$$\mathbf{w}_{k+1} = \mathbf{w}_k + \mu e_k \mathbf{x}_k - \rho \frac{\|\mathbf{w}_k\|_p^{1-p} \operatorname{sgn}(\mathbf{w}_k)}{\varepsilon + |\mathbf{w}_k|^{1-p}}, \quad (5)$$

where $\rho = \mu\gamma$ is an important parameter which weights the $p$-norm constraint and significantly affects the performance of the algorithm, $\varepsilon$ is a constant bounding the term and $\operatorname{sgn}(x)$ is the sign function, which is zero for $x = 0$, 1 for $x > 0$ and -1 for $x < 0$.

Compared to the standard LMS, the update of the $\ell_p$-LMS has an extra update term which attracts all the coefficients in the weight to zeros and thus, accelerate the convergence. However, it will lead to a slow down for the convergence of those taps that are optimized by the term $\mu e_k \mathbf{x}_k$ in the update equation, since it does not distinguish the different sparsity levels of the system [11].

### 2.2 Proposed $\ell_p$GC-LMS and $\ell_p$NGC-LMS algorithms

In order to conquer the above limitation of the $\ell_p$-LMS, inspired by [11], we introduce a gradient comparator for the $\ell_p$-LMS to obtain a new algorithm called the $\ell_p$GC-LMS, which selectively employs the zero attractor of $p$ norm on those taps that have the same polarity as the gradient of the squared instantaneous error. The update equation for the $\ell_p$GC-LMS is given by

$$\mathbf{w}_{k+1} = \mathbf{w}_k + \mu e_k \mathbf{x}_k - \rho \mathbf{G}_k \frac{\|\mathbf{w}_k\|_p^{1-p} \operatorname{sgn}(\mathbf{w}_k)}{\varepsilon + |\mathbf{w}_k|^{1-p}}, \quad (6)$$

where $\mathbf{G}_k = \operatorname{diag}[|\operatorname{sgn}(e_k \mathbf{x}_k) - \operatorname{sgn} \mathbf{w}_k|]/2$ is a diagonal matrix.

Furthermore, we also explored an improved version for the $\ell_p$GC-LMS, called $\ell_p$NGC-LMS, which employs a new gradient comparator that selectively zero-attract only taps that have the same polarity as the gradient of the mean squared error. Thus its update is obtained as

$$\mathbf{w}_{k+1} = \mathbf{w}_k + \mu e_k \mathbf{x}_k - \rho \mathbf{D}_k \frac{\|\mathbf{w}_k\|_p^{1-p} \operatorname{sgn}(\mathbf{w}_k)}{\varepsilon + |\mathbf{w}_k|^{1-p}}, \quad (7)$$

where $\mathbf{D}_k = \operatorname{sgn}\left(\frac{1}{S}\sum_{i=k-T}^{k} \mathbf{G}_i\right)$ is the "signed mean" version of the $\mathbf{G}_k$ above, and it should be emphasized that the integer $S$ (practically set to 5 or more) employed here is significant whose logic will be explained below.

The roles of gradient comparators $\mathbf{G}_k$ and $\mathbf{D}_k$ are elaborated as follows: Let $[\mathbf{w}]_i$ be the $i$th element of the real weight $\mathbf{w}$ of the unknown system, such that $[\mathbf{w}]_i > 0$, $[\mathbf{w}]_i < 0$ and $[\mathbf{w}]_i = 0$ are all the three possibilities for the range of value $[\mathbf{w}]_i$. Firstly, we consider the case $[\mathbf{w}]_i > 0$: (1) If $[\mathbf{w}_k]_i > [\mathbf{w}]_i$, then $E[e_k \mathbf{x}_k]_i = E[(\mathbf{w}^T \mathbf{x}_k + n_k - \mathbf{w}_k^T \mathbf{x}_k)\mathbf{x}_k]_i < 0$, and thus, $[\mathbf{G}_k]_{i,i} = 1$ with a high probability of at least 50% which depends on the signal noise ratio (SNR), whereas, $[\mathbf{D}_k]_{i,i} = 1$ holds up with much higher probability due to the employed trick of signed mean method, which takes the expectation to increase the likelihood of the equation $[\mathbf{D}_k]_{i,i} = 1$ in this case or $[\mathbf{D}_k]_{i,i} = 0$ below as much as possible. Practically we set the value of $S$ to be 5 or more to made it. (2) When $[\mathbf{w}_k]_i < [\mathbf{w}]_i$, if $[\mathbf{w}_k]_i < 0$, we have $[\mathbf{G}_k]_{i,i} = 1$ with a high probability and $[\mathbf{D}_k]_{i,i} = 1$ with a much higher probability. Similarly, we also have $[\mathbf{G}_k]_{i,i} = 0$ and $[\mathbf{D}_k]_{i,i} = 0$ with high and much higher probabilities respectively, if $[\mathbf{w}_k]_i > 0$. Secondly, the same logic reasoning will also apply to the case $[\mathbf{w}]_i < 0$. Finally, when $[\mathbf{w}]_i = 0$, we have $[\mathbf{w}_k]_i > 0$ or $[\mathbf{w}_k]_i < 0$, $[\mathbf{G}_k]_{i,i} = 1$ and $[\mathbf{D}_k]_{i,i} = 1$ with both high but different aforementioned probabilities are always concluded.

The $\ell_p$GC-LMS or $\ell_p$NGC-LMS algorithm works in the same way as the $\ell_p$-LMS when $[\mathbf{G}_k]_{i,i} = 1$ or $[\mathbf{D}_k]_{i,i} = 1$ holds up, i.e., the $p$ norm constraint is employed to attract the taps to zeros in this case; And they will be deduced to the standard LMS algorithm when $[\mathbf{G}_k]_{i,i} = 0$ and $[\mathbf{D}_k]_{i,i} = 0$, which neutralize the function of the zero attractor of the $\ell_p$-LMS. Additionally, it should be underlined that the global convergence and consistency of the $\ell_p$GC-LMS and $\ell_p$NGC-LMS remain problematic, which is inherited from the $\ell_p$-LMS whose cost function is not guaranteed to be convex. The Pseudo-codes for the $\ell_p$NGC-LMS algorithm is given in Table 1.

Furthermore, we can also derive some other gradient compared norm-constraint LMS algorithms including the $\ell_1$-LMS, $\ell_0$-LMS and some of their variants. Added to this, it is also worth mentioning that the variable step size (VSS) is known to have lower steady-state error as well as faster convergence. Thus, as a further extension, the VSS $\ell_p$GC-LMS and VSS $\ell_p$NGC-LMS can be easily derived in the same way as it has been added to the $p$-norm-like LMS in [10].

## 3 SIMULATIONS

Numerical simulations are carried out for several scenarios in this section to investigate the performances of the proposed $\ell_p$GC-LMS as well as the $\ell_p$NGC-LMS algorithm in terms of the steady-state mean square deviation (MSD, defined as $\mathrm{MSD}_k = \mathrm{E}[\|\mathbf{w} - \mathbf{w}_k\|_2^2]$) and convergence speed, and their results are compared with the $\ell_p$-LMS and the standard LMS algorithm in the settings of sparse system identification with different sparsity levels. Additionally, the superior performance of the $\ell_p$-LMS to other sparsity-aware modifications of LMS algorithms that are beyond the $p$ norm constraints, has been shown in Ref. [9], which, to be brief, is not detailed here, nor is the contrast with $\ell_1$GC-LMS in Ref. [11] as a modification of $\ell_1$-LMS, due to the parameter dependence problem in comparison with the $\ell_p$-LMS algorithm [9].

### 3.1 Example 1: sparse system with white Gaussian input

In the first experiment, we estimate a sparse unknown time-varying system of 16 taps with 1, 4 or 8 taps that are assumed to be nonzeros, making the sparsity ratio (SR) be 1/16, 4/16 or 8/16, respectively. The positions of nonzero taps are chosen randomly and the values are 1's or -1's randomly. Initially, we set SR = 1/16 for the first 500 iterations, and after that we have SR = 4/16 for the next 500 iterations, and then SR = 8/16 for the last 500 iterations which leaves a semi-sparse system. The input signal and observed noise are both assumed to be white Gaussian processes of length 516 with zero mean and variances 1 and 0.01, respectively, i.e., the signal noise ratio (SNR) is set to be 20 dB. Other parameters are carefully selected as listed in Table 2. Note that we use the same step size $\mu$ for all the four filters and the same $\rho$, $\varepsilon$ and $p$ for the three $p$ norm constraint LMS algorithms. All the simulations are averaged over 200 independent runs to smooth out their MSD curves.

Fig. 1 shows the MSD curves of the proposed $\ell_p$GC-LMS and $\ell_p$NGC-LMS algorithms with respect to the number of iterations with different sparsity levels, i.e., SR = 1/16, 4/16 and 8/16 for the three iterative stages, compared with the standard LMS and $\ell_p$-LMS algorithms. As shown in Fig. 1, we can see that in the very sparse case (SR=1/16) for the first 500 iterations, the $\ell_p$NGC-LMS achieves the best performance in terms of convergence speed and stable error, while the $\ell_p$GC-LMS works a little worse than the $\ell_p$-LMS, probably due to the aforementioned probability problem of $\mathbf{G}_k$, It still perform better than the standard LMS. With the sparsity decreasing (SR=4/16, 8/16 for next 2 stages with 500 iterations each), the performances of all the three $p$-norm constraint LMS algorithms deteriorate as expected. However, the $\ell_p$NGC-LMS still has better performance than the $\ell_p$-LMS and the $\ell_p$GC-LMS, and the $\ell_p$GC-LMS still performs better than the $\ell_p$-LMS in these two cases when the system is semi-sparse. Thus, it can be concluded that the proposed $\ell_p$GC-LMS and $\ell_p$NGC-LMS algorithms outperform the $\ell_p$-LMS for a system with different sparsity levels and white Gaussian input.

Table 1. Pseudo-codes of the $\ell_p$NGC-LMS Algorithm

| Given | $\mu$, $\rho$, $N$, $S$, $\varepsilon$, $\mathbf{w}$, $\mathbf{x}$, $L$ |
|---|---|
| Initial | $\mathbf{w}_0$=zeros($N$,1), $p$, $\mathbf{Y}$=zeros(1,$L$) |
| for | $i = 1,2,...,(L\text{-}N+1)$ |
| | $y_k = \mathbf{w}^T \mathbf{x}_k + n_k$ |
| | $e_k = y_k - \mathbf{w}_k^T \mathbf{x}_k$ |
| | $\mathbf{w}_{k+1} = \mathbf{w}_k + \mu e_k \mathbf{x}_k - \rho_p \mathbf{D}_k \dfrac{\|\mathbf{w}_k\|_p^{1-p} \mathrm{sgn}(\mathbf{w}_k)}{\varepsilon_p + |\mathbf{w}_k|^{1-p}}$ |
| | $\mathbf{G}_k = \mathrm{diag}[|\mathrm{sgn}(e_k \mathbf{x}_k) - \mathrm{sgn}\,\mathbf{w}_k|]/2$ |
| end | $\mathbf{D}_k = \mathrm{sgn}\left(\dfrac{1}{S}\sum_{i=k-T}^{k}\mathbf{G}_i\right)$ |

Table 2. Parameters of the algorithms in the example 1.

| Algorithms | $\mu$ | $\rho$ | $\varepsilon$ | $p$ | $S$ |
|---|---|---|---|---|---|
| LMS [1] | 0.05 | / | / | / | / |
| $\ell_p$-LMS [9] | | 0.0008[a] | 0.05 | 0.5[①] | / |
| $\ell_p$GC-LMS | | 0.0003[b] | | | |
| $\ell_p$NGC-LMS | | 0.0001[c] | | | 5 |

[a]~[c] are for SR=1/16, 4/16 and 8/16, respectively.

Fig.2 shows the MSD curves of these algorithms tested with respect to different sparsity levels (with fixed $\rho$ = 0.0005), from which we can see that the MSDs of all the three $p$-norm constraint LMS algorithms increase with the SR, whereas the MSD of the standard LMS is stable. As observed from Fig.1, the $\ell_p$NGC-LMS has the lowest MSD when the system is sparse, then comes by the $\ell_p$GC-LMS. It is important to note that the $\ell_p$NGC-LMS has lower MSD than the $\ell_p$-LMS whatever the sparsity level of the system is, under the common settings in this simulation.

---

[①] Generally, 0.5 is the default value for $p$ in most $\ell_p$-LMS related problems with fixed $p$ values, whereas we have explored and achieved a new $\ell_p$-LMS with different methods of variable $p$ and better performances, which will be specified in the following paper, due to the space limitations.

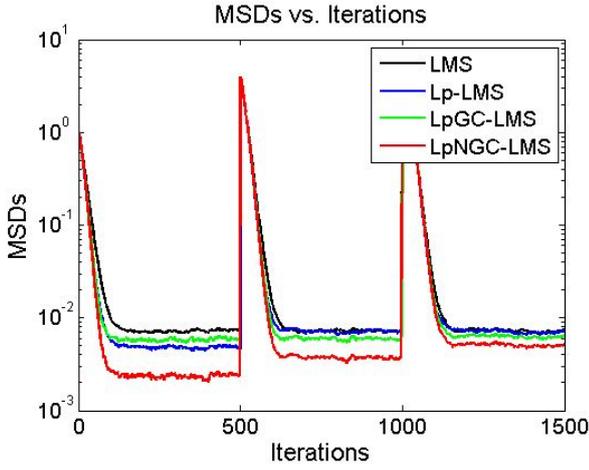

Fig. 1. MSD curves of different algorithms with SNR = 20 dB and SR=1/16, 4/16 and 8/16, respectively.(white Gaussian input)

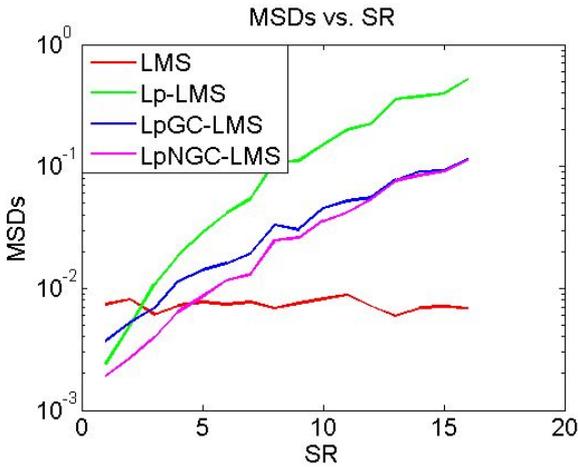

Fig. 2. MSD curves of tested algorithms with fixed $\rho = 0.0005$, different sparsity levels and SNR = 20 dB.

### 3.2 Example 2: sparse system with correlated input

The unknown system in the second example is the same as the previous one, except that we change the switching times for different sparsity levels to the 3000$th$ iteration and the 6000$th$ iteration, respectively. The input is now a correlated signal which is generated by $x_{k+1} = 0.8x_k + u_k$ and then normalized to variance 1, where $u_k$ is a white Gaussian noise with variance $10^{-2}$. And the variance of the observed noise is set to $10^{-1}$ in this example. Other parameter choices for all the algorithms tested are listed in Table 3. As it is set in the first example, we choose the same $\mu$, $\rho$, $\varepsilon$ and $p$ for all the tested algorithms if they are required. However, to solve the problem underlined in the previous example that the $p$ norm constraint LMS algorithms may preform worse than the standard LMS due to the choices of the parameter, different $p$-norm constraint weight $\rho$'s are selected for different sparsity levels in this case, which would utilize the sparse constraints better, and thus achieve lower steady-state MSD values than that of the standard LMS algorithm. In addition, all the MSD curves of the simulations are smoothed via 200 Monte-Carlo runs as well.

Table 3. Parameters of the algorithms in the example 2.

| Algorithms | $\mu$ | $\rho$ | $\varepsilon$ | $p$ | $S$ |
|---|---|---|---|---|---|
| LMS [1] |  | / | / | / | / |
| $\ell_p$-LMS [9] | 0.015 | 0.0005[a] | 0.1 | 0.5 | / |
| $\ell_p$GC-LMS |  | 0.00005[b] |  |  |  |
| $\ell_p$NGC-LMS |  | 0.00001[c] |  |  | 5 |
| [a]~[c] are for SR=1/16, 4/16 and 8/16, respectively. |

Fig. 3 shows the MSD curves of the algorithms tested for different sparsity levels, i.e., SR = 1/6 for the first 3000 iterations, SR = 4/16 for the 3001$th$ to 6000$th$ iteration and SR=8/16 for the next 3000 iterations left. It can be seen from Fig. 3 that similar performance trends are observed as in the first example, i.e., most of the observations from Fig. 1 also hold true for this case where the input signal is related. For systems of different sparsity levels, the $\ell_p$NGC-LMS and $\ell_p$GC-LMS always achieve the faster convergence and lower steady-state MSD than the $\ell_p$-LMS. Moreover, the $\ell_p$NGC-LMS always performs better than the $\ell_p$GC-LMS and then followed by the $\ell_p$-LMS in any case. However, as the sparsity ratio increases, the behaviors of all the $p$ norm constraint LMS algorithms tested in this paper which are expected to be better than the standard LMS algorithm, depend strictly on the parameters $\rho$ and the SNR of the input. Specifically, as SR increases from 1/16 to 8/16, the smaller $\rho$, and / or the smaller SNR in a certain range, the better performances which the $p$-norm constraint LMS algorithms achieve, and the smaller performance gaps between the $\ell_p$NGC-LMS and $\ell_p$GC-LMS, the $\ell_p$GC-LMS and $\ell_p$-LMS. Anyway, the $\ell_p$NGC-LMS performs the best among these three $p$-norm constraint LMS algorithms, and better than or equivalently to the standard LMS if proper parameters are selected as in the settings of this example.

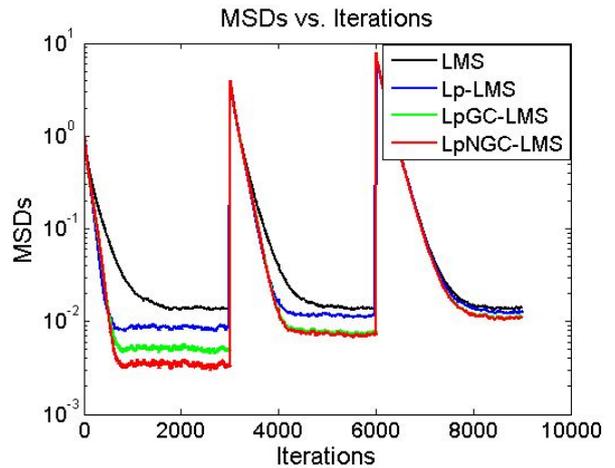

Fig. 3. MSD curves of different algorithms with SR=1/16 ($\rho = 5\times10^{-4}$), 4/16 ($\rho = 5\times10^{-5}$) and 8/16 ($\rho = 10^{-5}$), respectively. (correlated input)

### 3.3 Example 3: system of ECG-like impulse response

In this example, we estimate a sparse system with an ECG-like impulse response (IR) of 256 taps, in which 28 of them are nonzeros. Fig. 4 shows the tap vector of the system to be identified, which is sampled from an ECG signal and then processed into a simpler version with most of the small taps set to zeros, to make it sparse for our simulations. The input signal and additional noise are white Gaussian processes with variance 1 and 0.1, respectively. Other parameters are selected in Table 4 and all the simulations are performed 200 times. Note that we choose these parameters that are different from the previous two examples to yield better performance in term of convergence speed and steady-state error plotted in MSD curves, which is shown in Fig. 5.

Table 4. Parameters of the algorithms in the example 3.

| Algorithms | $\mu$ | $\rho$ | $\varepsilon$ | $p$ | $S$ |
|---|---|---|---|---|---|
| LMS | 0.005 | / | / | / | / |
| $\ell_p$-LMS | | 0.000007 | 0.1 | 0.5 | |
| $\ell_p$GC-LMS | | | | | |
| $\ell_p$NGC-LMS | | | | | 5 |

From Fig. 5, one can see that the previous observations still hold up in this much longer sparse system, the proposed $\ell_p$NGC-LMS and $\ell_p$GC-LMS outperform the $\ell_p$-LMS and standard LMS algorithm with faster convergence speed and lower steady-state MSD.

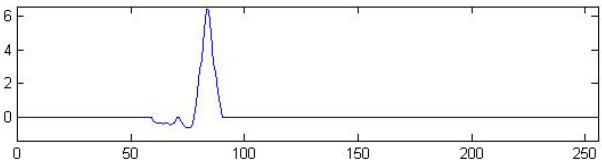

Fig. 4. The impulse response of the system in the example 3.

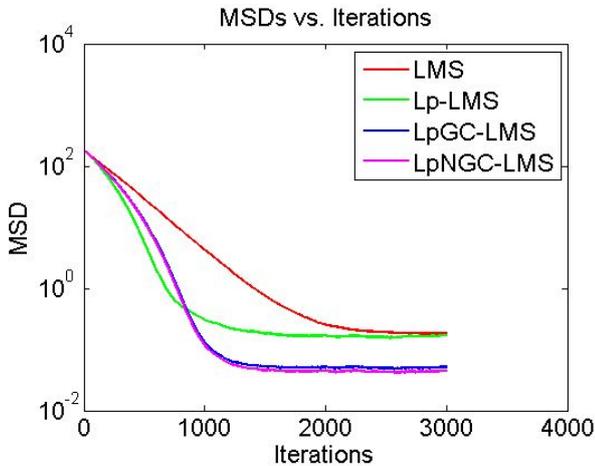

Fig. 5. MSD curves of different algorithms with SNR = 10 dB and SR=28/256. (ECG-like system)

## 4 CONCLUSION

We propose two gradient compared *p*-norm penalty LMS algorithms, i.e., the $\ell_p$GC-LMS and $\ell_p$NGC-LMS algorithm, as extensions of the conventional $\ell_p$-LMS algorithm which is established for sparse adaptive filtering recently. Two gradient comparators are employed to selectively apply the zero attractor in $\ell_p$-LMS for only those taps that have the same polarity as that of the gradient of the squared instantaneous error or mean squared error. Numerical simulation results show that the proposed algorithms yield better performances than those of the standard LMS and $\ell_p$-LMS algorithms in sparse system identification settings in term of convergence speed and steady-state mean square deviation.

Summarily, we can conclude from the above statements in this paper that the proposed $\ell_p$GC-LMS and $\ell_p$NGC-LMS are actually two approaches to weight the *p*-norm constraint for the $\ell_p$-LMS algorithm to select certain taps in order to perform better, and there are also some other methods to achieve this. Therefore, our future work will still focus on the choices of the parameters of the $\ell_p$-LMS algorithm for sparse system identification, including exploring the relationship among the weight of norm constraint, the sparsity ratio of the impulse response and the signal-noise ratio of the input, considering an adaptive parameter *p* and *ρ* for different sparsity levels and / or different SNR, and developing new $\ell_p$-LMS algorithms for better identifying an unknown system which is not assumed to be sparse or non-sparse, etc.